\documentclass[conference]{IEEEtran}
\IEEEoverridecommandlockouts
\usepackage{cite}
\usepackage{amsmath,amssymb,amsfonts}
\usepackage{algorithmic}
\usepackage{graphicx}
\usepackage{textcomp}
\usepackage{xcolor}
\def\BibTeX{{\rm B\kern-.05em{\sc i\kern-.025em b}\kern-.08em
    T\kern-.1667em\lower.7ex\hbox{E}\kern-.125emX}}
\begin{document}

\title{Writing summary for the state-of-the-art methods for big data clustering in distributed environment\\
\thanks{Identify applicable funding agency here. If none, delete this.}
}

\author{\IEEEauthorblockN{Dipesh Gyawali}
\IEEEauthorblockA{\textit{Department of Computer Science and Engineering} \\
\textit{Louisiana State University}\\
Baton Rouge, United States \\
dgyawa1@lsu.edu}
}

\maketitle

\begin{abstract}
Big Data processing systems handle huge unstructured and structured data to store, process, and analyze through cluster analysis which helps in identifying unseen patterns to find the relationships between them. Clustering analysis over the shared machines in big data technologies helps in deriving the relations and making decisions using data in context. It can handle every form of raw, tabular data along with structured, semi-structured, and unstructured data. The data doesn't have to possess linearity property. It can reflect associative and correlative patterns and groupings. The main contribution and findings of this paper are to gather and summarize the recent big data clustering techniques, and their strengths, and weaknesses in any distributed environment. 
\end{abstract}


\section{Introduction}
Big Data technologies process huge data to derive meaningful results from them. Data in the real world could be of different formats like text, images, graphs, audio, etc. These data are present in unstructured, semi-structured, and structured forms. Real-world data is in raw form and doesn't contain meaningful information. When processed and analyzed, meaningful information can be extracted. Cluster analysis helps in identifying the hidden patterns and relationships in data that leverage a lot of power to experiment and discover new results. This paper discusses the primary ideas of five research papers in the area of shared cluster architecture in a distributed environment. Most of these papers elaborate upon how the performance of shared clustering in big data can be enhanced by introducing new methodologies and approaches within shared clusters. These papers also talk about how scaling can be done in this architecture without facing any interruptions. The strengths and weaknesses of each clustering technique are also discussed briefly. 

\section{Summary}

\subsection{Multi-objective Fuzzy-Swarm Optimizer for Data Partitioning}
Data partitioning is extremely useful in today's context when we try to use big data applications in distributed environments. While handling huge datasets, it is significant to make data partition scalable and elastic. Multi-objective fuzzy swarm optimizer is based on clustering for data partitioning, data reduction, and processing. In traditional technologies like Apache Hadoop and Spark, shared-nothing architecture is used as every node inside them is free and independent of other frameworks within both data and resources. In the Hadoop framework, if we want to combine several machines into a computing cluster, it will be more expensive in terms of investment and resources and it is sometimes infeasible because of its unavailability. This method helps to reduce the expense of computing clusters and uplift the performance of big data processing in computing by acquiring high throughput and low latency. 

In this paper, the optimization using multi-objective functions is introduced with the application of swarm optimization and fuzzy-logic. Complex datasets require multi-objective optimization which contains multiple constraints for the best possible solutions. The stopping criteria are extremely important to get an optimized solution which is chosen as a total number of maximum iterations for fuzzy-swarm optimization. This algorithm is used to optimize parameters for maximizing the given objective function and reducing the time complexity at the same time. After the whole iteration is completed, the optimized algorithm helps in identifying the best search space in the clusters in addition with good fitness. As compared to previous traditional approaches which solely focus on designing only a single objective and multi-objective clustering algorithms to partition data, this experiment integrates some solutions with the pre-existed designs to improve the efficacy. 

With the increase in data and users, various types of data are collected and analyzed with different data mining tools. Categorization of data into various clusters that are collected from multiple sources are done. This paper has proven the effectiveness of swarm optimization for partitioning the data into various clusters for analyzing and other operations. It improves the global search ability for the best solution by using a multi-objective optimization approach and gains higher accuracy compared to single-objective swarm optimization algorithms in the context of complex data.

\subsection{CHIME: A checkpoint-based approach in order to improve the performance of the shared clusters}
Tasks can come up with higher and lower priorities according to their need in huge commercial cloud platforms utilizing big data technologies. For handling the task, a preemptive algorithm should be adopted in that case. It is obvious that higher priority tasks are executed immediately as compared to lower priority tasks in big data cloud platforms like Microsoft Azure, GCP, and Amazon Web Services. This is helpful when there are not enough resources to perform cloud operations at a certain period of time. And, when the resources are available later, the task is re-executed. However, these low-priority tasks could be restarted many times during peak times because of the lack of resources which will induce memory-related problems across hard disk drives and CPU cores. To handle this, checkpoint technology is introduced and adopted which generally checkpoints the low-priority tasks that have been restarted periodically. This paper talks about the checkpoint technology to improve the performance on clouds by handling low-priority tasks. 

This paper has proposed a method called 'CHIME' to handle task handling with different priorities that utilize checkpoint technology and introduce a waiting list approach. Generally, low-priority tasks are sacrificed in order to perform high-priority tasks when the system is under a heavy workload and there is a lack of resources. Microsoft once reported that 21\% of its tasks were killed due to preemptive scheduling in its Dyrad cluster. For this, CHIME has been one of the successful implementations in handling task scheduling on the shared clusters.

When the priority of one task is greater than the priority of tasks that is already being executed with optimum machine performance, those tasks will be kept on a combined waiting list and the previous task will be executed. The number of checkpoints performed is also decreased by setting a checkpoint checking condition. Performing unnecessary overhead will add redundant overhead to the cluster. Checkpointing is only done when the occupancy of the entire system decreases more than the threshold C. When occupancy becomes greater than C, checkpointing will be started for low-priority tasks. This approach starts the task in the combined waiting list as soon as the resource starts becoming free. This combined waiting list can hold all the tasks including tasks never checkpointed, previously checkpointed, and were killed without any checkpoint.

The introduction of CHIME has improved the scheduling techniques of tasks being executed on big data platforms which ultimately improves cloud performance, and helps in proper resource utilization for data operations and handling. 

\subsection{CoBell: Runtime Prediction for Distributed Dataflow Jobs in Shared Clusters}
When developing big data applications, users/programmers don't know much about the distributed part that big data frameworks generally handle. Users can easily develop huge applications in a parallel computing environment using big data technologies. They can assign resources for the workloads. They don't fully understand resource utilization and system dynamics. Some technologies are developed to select the amount of resources required for fulling the tasks. This paper discusses 'CoBell' which is a resource allocation system that aims to enhance the runtime prediction of multiple jobs in shared clusters. It reserves the required resources for the users based on runtime prediction.

CoBell has been implemented as a job submission tool for YARN. Its main goal is to enhance the runtime prediction's accuracy for interfered workloads by storing knowledge for the co-located jobs. Users submit jobs along with their constraints (scale-out) and runtime to CoBell which is later converted into resource reservation. It is achieved from the record of already executed jobs. After the submission of a new job, it queries its database to find out the runs of the previous job. These old runs are utilized to build the prediction model. Its runtime model helps in accumulating the knowledge about the co-located jobs to improve runtime performance. Two types of models are provided - a strong model which performs extrapolation, and another non-parametric model that provides an alternative to interpolation. After building the model, it is then used for predicting the runtime resources. 

\subsection{Enabling Elastic Stream Processing in Shared Clusters}
Various big data frameworks like Hadoop MapReduce helps in handling large data sets, process, storing, and analyzing them, Recently, there has been increasing in the number of applications that require computing real-time data being generated like dynamic content delivery, intrusion detection systems, graph processing, etc. Nowadays, other big data frameworks like Spark Streaming and Storm support live data processing that runs on shared clusters. Here, low cost needs to be achieved to maintain the proper cluster performance which is one of the challenges in stream data processing. 

Due to workloads and requirements of system resources, the velocity of inputs in data stream processing systems is not constant. It's not possible for current system to scale up the resources dynamically without any service interruptions. For scaling up, the application first needs to be shut down, reconfigured, and restarted. This is extremely unreliable where real-time data processing is required due to network interruptions. In addition to this, there arises a problem with the application state that results in work lost and ambiguous results. Elastic data streaming helps to add up resources dynamically without the requirement of system shutdown. This is achieved by the adoption of the stateful scaling mechanism and cost-dependent scaling. The researchers added a Hadoop YARN cluster resource manager to coordinate resources between batch and stream processing. 

The elastic data processing also helps in saving the system's state which is sent to the global state manager component. The global manager component helps to coordinate the transfer of the operations at the place of scaling operations. Congestion Detection Monitor (CDM) is used to detect congestion. It keeps track of the CPU performance, and memory and analyzes congestion when the threshold of machines is crossed. If any of the resources in the system is bottle-necked, CDM will increase the number of workers in the topology. Likewise, the workers are decreased if no system metrics are overloaded. 

\subsection{Building Strongly Reliable and High-Performance Storage Cluster with Network and Attached Storage Disks Using PCs}
The authors of this paper has discussed a prototype implementation of a storage cluster where every node is equipped with network-attached storage disks forming a Network Attached Cluster (NAS). Memory is shared between nodes inside the clusters to implement a strong cache to disk policy. Nowadays, the users can create a storage cluster using off-the-shelf commodity parts in the presence of tremendous power. Every PC node is attached to local storage devices. These devices form the Storage Area Network.

Users/Clients can use distributed system protocols like Common Internet File System (CIFS) and Network File System (NFS). Basic file operations command like create, read, write and NAS clusters can also be used. The system is server-less where nodes are responsible for storing the file data and managing metadata. The files received are divided across the nodes for parallel processing. Global Unified Content Cache (GUCC) utilizes high-level abstraction for dealing with global caching, and remote resources. When the user requests are missed by local caches in each node, they are handled by the global cache. The daemon in the cluster includes functions like managing the global cache and providing files for nodes across the clusters. 

Daemon checks the usage of the global cache periodically to provide a reasonable cache policy. The daemon will move the data blocks from the remote to balance the node load when a particular node’s capacity goes below a threshold value. In a nutshell, this paper is based on the notion of GUCC which is different than the traditional caching that lets the cache miss of one node be handled by a different remote node. This increases the overall memory of the shared cluster to go beyond the local working limit.

\section{Pros and Cons}

\subsection{Multi-objective Fuzzy-Swarm Optimizer for Data Partitioning}\label{AA}
There is an improvement in efficiency for data partitioning as compared to the previous approach as it uses the solution from other sources to integrate in their design. Using a multi-objective optimizer, the data partitioning process is more reliable and accurate comparitively to a single objective optimizer.

When trying to use a multi-objective optimization function, it's sometimes difficult to converge them simultaneously. The optimization directly depends upon the type of optimizer used to determine how quickly and efficiently the data partition process takes place.

\subsection{CHIME: A checkpoint-based approach in order to improve the performance of the shared clusters}
The choice of preemption policy is only applied when absolutely necessary in CHIME. During the peak time of clusters, restoring and preemptive operations take place. This helps in preventing the loss of low-priority tasks by developing a proper mechanism for handling all the tasks in order using a combined waiting list. This is most effective in real-time environment when the decision should be made instantaneously.
Regarding the cons, sometimes the execution of high-priority tasks could be delayed due to the adoption of the FIFO method for a combined waiting list. In FIFO, the task is not executed based on priorities. Rather they are treated equally.

\subsection{CoBell: Runtime Prediction for Distributed Dataflow Jobs in Shared Clusters}
CoBell depends on the job submissions and the scale-out constraints with the job history from database to detect the runtime environment that is easily accessible in the system. Hence, it doesn't require many inputs or parameters to detect the environment. Dynamic scaling in or out is not required since the resources are already reserved before. It also utilizes the knowledge of concurrently running jobs to predict the required resources in the shared clusters. 
Sometimes, there is a greater number of jobs in real-world applications. When there are big applications with a random number of job interference, this approach of runtime predictions could be an issue.

\subsection{Enabling Elastic Stream Processing in Shared Clusters}
The main contribution of this approach is to remove the requirement of system shutdown and reconfiguration like that of traditional scaling approaches which is one of the main advantages. The absence of frequent shutdowns helps to prevent data loss that directly impacts network interruptions. The automatic congestion detection system enables limited user monitoring and manual intervention. The elastic stream processing helps to identify CPU bottlenecks. The dynamic load-aware scheduling mechanism can elastically scale out or scale down components according to the different requirements. This scaling mechanism helps to save application state to prevent the loss of work and prevent performance deterioration. 
The model is based on rule-based assumptions which make it less flexible. It generally checks the threshold levels of the CPU. Instead of a rule-based model, the predictive model could be better. Elastic stream processing generally doesn't handle the processing of a single machine with multiple clusters which are common among various applications. 

\subsection{Building Strongly Reliable and High-Performance Storage Cluster With Network Attached Storage Disks Using PCs}
There are no limits on the number of nodes present inside the cluster which makes this approach scalable. Two types of caching make the data transfer faster. The global cache mechanism helps to access data when the local cache is unable to access it. One node's memory can be utilized by others during data transfer which reduces memory overhead. There is an easy switching mechanism between RAID1 and RAID5 schemes to achieve the highest performance. GUCC solves the disk latency issues. This approach is also suitable for small and large writes.
This system sometimes doesn't support multiple operating systems as it is specifically designed for Linux. The overall system's performance is achieved from the hardware's optimum performance and speed which is not always the case. This introduces the biasness in the performance from this context. Moreover, Network Attached Storage is highly dependent on commodity machines and chips.

\section{Conclusion}
Different state-of-the-art approaches for shared clustering architecture have been studied in the paper that is prevailing in the distributed environment. Cluster architecture helps to make systems more scalable and reliable that could handle big data operations smoothly. Still, there are many approaches that are not defined in this paper that work well in distributed clusters. The summary, strengths and, weaknesses of the five approaches are well defined in the paper.

\vspace{12pt}

\end{document}